\documentclass[useAMS,usenatbib]{mn2e}
\usepackage{graphicx}

\title[PWV above ORM]{The Infrared Astronomical Characteristics of Roque de los Muchachos Observatory: precipitable water vapor statistics}

\author[Garc\'{\i}a-Lorenzo et al.]{B. Garc\'{\i}a-Lorenzo$^{1,2}$\thanks{E-mail:
bgarcia@iac.es}, A. Eff-Darwich$^{3}$, J. Castro-Almaz\'an$^{1}$, N. Pinilla-Alonso$^{4}$, C. Mu\~noz-Tu\~n\'on$^{1,2}$, \& Rodr\'{\i}guez-Espinosa, J.M.$^{1,2}$\\
1 Instituto de Astrof\'{\i}sica de Canarias, C/Via Lactea S/N, 38305-La Laguna, Tenerife, Spain \\
2 Dept. Astrof\'{\i}sica, Universidad de La Laguna, C/ Astrof\'{\i}sico Francisco S\'anchez, E-38205 Tenerife, Spain \\
3 Dept. Edafolog\'{\i}a y Geolog\'{\i}a, Universidad de La Laguna, C/ Astrof\'{\i}sico Francisco S\'anchez, E-38205 Tenerife, Spain \\
4 NASA Ames Research Center, MS 245-3, Moffet Field, 94035-1000, CA}

\begin{document}

\date{Accepted ..... Received .....; in original form .....}

\pagerange{\pageref{firstpage}--\pageref{lastpage}} \pubyear{2004}

\maketitle

\label{firstpage}

\begin{abstract} 

The atmospheric water vapor content above the Roque de los Muchachos Observatory (ORM) obtained from Global Positioning Systems (GPS) is presented. GPS measurements have been evaluated by comparison with 940nm-radiometer observations. Statistical analysis of GPS measurements points to ORM as an observing site with suitable conditions for infrared (IR) observations, with a median column of precipitable water vapor (PWV) of 3.8 mm. PWV presents a clear seasonal behavior, being Winter and Spring the best seasons for IR observations. The percentage of nighttime showing PWV values smaller than 3 mm is over 60\% in February, March and April. We have also estimated the temporal variability of water vapor content at the ORM. A summary of PWV statistical results at different astronomical sites is presented, recalling that these values are not directly comparable as a result of the differences in the techniques used to recorded the data.

\end{abstract}


\section{Introduction}

The triatomic molecules H$_2$0, CO$_2$ and O$_3$ are the main responsible for reducing the few transparent windows on
 the infrared atmospheric transmission spectrum, and producing absorption bands that are difficult to correct during
 the processing of astronomical data. Out of these windows, the atmosphere is opaque and observations at those wavelengths
 have to be carried out from space. Sites with large atmospheric water vapor levels have narrow and unstable transparent windows. High levels of water
vapor reduce the atmospheric transparency, but also increase the thermal infrared background.

 It is assumed among the astronomical community that sites placed at lower altitudes
 are less suitable for high-quality IR observations; however, other parameters such the troposphere thickness might be also playing an
important role \citep{2004SPIE.5572..384G}. Indeed, the evaluation of the IR quality
 at the Observatorio del Teide in the Canary Islands, Spain \citep{1985A&A...150..281M} and the comparison
 with prediction models \citep{1992AJ....104.1650C} indicate that El Teide site, at $\sim$2400 m above sea level, might be
as good as Mauna Kea (4200 m) for IR astronomy in the 1-5 $\mu$m window \citep{1998NewAR..42..533H}. Observational results
pointing to a similar conclusion were reported for Roque de los Muchachos observatory (ORM) on the island
of La Palma (Canary Islands, Spain), also at $\sim$2400 m above sea level, when comparing infrared observations from the
3.58m Telecopio Nazionale Galileo with data obtained at the 10m Keck telescope on Mauna Kea astronomical
observatory (Hawaii, USA) (http://www.tng.iac.es/news/2003/03/21/nics$\_$refurbish).

There are many parameters accounting for the quality of an astronomical site, namely seeing, cloud cover, ground winds,
high-altitude winds, etc (see \cite{2007RMxAC..31...36M} for a review on the characterization of these parameters at ORM). The water
vapor content is an important parameter affecting the IR quality of astronomical sites. Conditions for IR astronomical observations were classified in terms of precipitable water vapor (PWV hereafter) in four divisions \citep{1998NewAR..42..537K}: 1) Good or excellent $\rightarrow$ PWV $\leq$ 3 mm; 2) Fair or mediocre $\rightarrow$ 3 $<$ PWV $\leq$ 6 mm; 3) Poor $\rightarrow$ 6 $<$ PWV $\leq$ 10 mm; and (4) Extremely poor $\rightarrow$ PWV $\geq$ 10 mm . The development of IR
instrumentation and the requirements for current large and future extremely large telescopes demand a proper
characterization of PWV  and statistical studies of large temporal databases (covering years). The fraction of
nights with good IR conditions (small column of water vapor) as a function of the epoch of the year will allow an optimal scheduling
of telescope observing time. 
The total atmospheric water vapor contained in a vertical column of unit cross-sectional area extending between any two
specified levels is known as precipitable water vapor and it is commonly expressed in terms of the height to which that
water substance would stand if completely condensed and collected in a vessel of the same unit cross section \citep{AMS00}.
 PWV is also referred to as the total column water vapor \citep{ferrare02}.
Measurements of PWV can be obtained in a number of ways, from {\it in situ} measurements (radiosondes) to remote sensing
techniques (photometers, radiometers, GPS, Imaging Spectroradiometers on satellites, etc). Radiosondes have been the
primary {\it in situ} observing system for monitoring water vapor; however, the operability of radiosondes is limited due to
 running costs and decreasing sensor performance in cold dry conditions \citep{2003JGRD..108.4651L}. Usually, radiosondes are expected to measure
PWV with an uncertainty of a few millimetres, which is considered to be the standard accuracy of PWV for
meteorologists \citep{2001JAtOT..18..830N}. Near infrared (NIR) radiometers measuring methodology assumes plane-parallel atmosphere,
hence it is only satisfactory for measurements near the Zenith. Moreover, their uses are limited to photometric and bright
 nights (from first quarter to last quarter Moon). Errors of 20\% or below are estimated for PWV measurements using NIR
 radiometers in the range 3-10 $\mu$m, whereas errors in the 0.5-1$\mu$m range might be as high as 40\% \citep{1989PASP..101..441Q}.
Microwave water vapor radiometers (WVR) observations are not reliable when liquid water is present on the WVR frequency window \citep{2001JApMe..40....5L}
 or when there is significant scattering from liquid water droplets and ice crystals in the field-of-view \citep{Zhang99}.

GPS is an increasingly useful tool for measuring PWV, which has gained a lot of
 attention in the meteorological community. The GPS procedure to estimate the PWV is based on the fact that the propagation of electromagnetic waves through the atmosphere
 is drastically affected by variations on the
 refraction index of the troposphere, which depends on the water vapor pressure, air pressure, and
 temperature. The consequence is an induced delay in any signal crossing the atmosphere, being due to a
 combination of a hydrostatic and a water vapor delay. The hydrostatic delay is very stable and has a direct
 relationship with local atmospheric pressure. The second component, the wet delay, is directly related to the water
vapor content in the atmosphere above the site where the measurements are taken. The rapid temporal variations of the water vapor affects its prediction and proper measurement. The overall tropospheric delay at a GPS
 station allows the estimation of the PWV with a high-degree of accurancy \citep{1994JApMe..33..379B, Boco06, 2009JGeod..83..537J}, although under very dry environments GPS might underestimate the PWV content \citep{schneider09}. GPS
 system provides a better spatial coverage and continuous PWV estimations in comparison with other techniques 
\citep{2000GeoRL..27.1915G}. The comparison of GPS, radiosondes and WVR \citep{2000ITGRS..38..324E, 2001JAtOT..18..830N, 2003JGRD..108.4651L} shows agreement generally at the level of 1-2 mm of PWV, corresponding to 7-13 mm of zenith wet delay (error $<<$ 10\%).

In this paper, we present statistical results on the PWV above the Roque de los Muchachos Observatory (La Palma, Spain) derived from GPS measurements for the period spanning from June 2001 to December 2008. For comparison purposes, we also present PWV statistical results for Mauna Kea site derived also from GPS data.

\section{DATA SOURCES}
\subsection{PWV measurements at ORM}
The Roque de los Muchachos Observatory is located at an altitude of 2396 m above sea level on the island of La Palma 
(Canary Islands, Spain) at latitude 28$^0$ 45$´$ North and longitude 17$^0$ 53$´$ West. This observatory is one of the final
candidate sites for the location of the 42 m European Extremely Large Telescope. 
 The Spanish Instituto Geogr\'afico Nacional has installed
 permanently since May 2001 a GPS receiver at ORM. The GPS antenna is placed on a very stable
 monument which ensures that the GPS satellite data recorded at the station does not contain any spurious antenna
 movement that could affect the scientific exploitation of the GPS data. This permanent GPS is part of the international
 network -- {\it European Reference Organisation for Quality Assured Breast Screening and Diagnostic Services} (www.euref.eu) 
-- of GPS stations \citep{2004AGUFM.G53A0115K}.
 The data from the GPS station at the ORM for the period spanning from June 2001 to December 2008 were processed by Soluciones
 Avanzadas Canarias S.L. (www.sacsl.es), a company devoted to GPS data processing
 for the International Global Navigation Satellite Systems Service \citep{2003AdSpR..31.1911R}. Global GPS data were processed
 using the dual frequency data recorded at around 35 stations worldwide including the station at ORM. The
 daily data processing runs take data from each station using all the data available above 10 degrees
 of elevation at each ground station every 5 minutes (epoch). The satellite orbits are considered fixed as downloaded from the International Global Navigation Satellite Systems Service (www.igs.org).  The Tropospheric Zenith Delay (TZD) for each station is estimated every two hours with an initial value of 2.1 meters and an a-priori sigma of 2 meters, hence being considered ``free''. The long standing stations have coordinates/velocities published in the International Terrestrial Reference Frame \citep{2007JGRB..11209401A} which are very precise and can be kept almost fixed. The coordinates for the stations in study are estimated using the initial conditions with a 3 mm a-priori sigma. This allows the coordinate estimation to be adjusted slightly over the processing arc to accommodate small daily variations from the published International terrestrial reference frame positions.

In each observation epoch, each station has measurements to all the satellites in-view. Nominally each station records measurements to 8-12 satellites per epoch. This means that for the TZD estimation (one value every two hours per station) each TZD value is averaged over 24 epochs (5-minutes each). Therefore, more than $\sim 200$ individual measurements (24 epochs $\times$ 8-12 satellites) contribute to the TZD estimation. A two-hour averaging time is standard in GPS processing for TZD, although shorter estimations can be performed. More details on the theoretical basis for the ground-based GPS technique, the background physics and potential applications can be found in e.g. \cite{1992JGR....9715787B}. The accuracy of GPS measurements are not limited by the system receiver noise, but largely by modeling errors. The uncertainties of the system are multidimensional, including errors from satellite orbits, errors in the mapping functions used for the atmosphere delay estimations, scattering in the vicinity of the GPS antenna, the effect of any radome in the antenna, etc. An introduction and description of GPS data processing can be found, for example, in \cite{blewitt98} or \cite{leick04}.  

The TZD derived from the GPS data processing is composed of two contributions,
 the hydrostatic (ZHD) and the wet (ZWD) delays. The ZHD can be calculated through the latitude ($\phi$ in degrees), altitude
 (H in meters) of the GPS station \citep{1973BGNS..107...13S, 1972GMS....15..247S} and the atmospheric pressure (P in hPa) according 
to the following expression:
\begin{equation}
                    ZHD = \frac{22.768\times10^{-4}\times P}{1 - 26.6\times10^{-4} cos(2\phi) - 0.28\times10^{-6}\times H}
\label{eq1}
\end{equation}

The ZWD accounts for the PWV in the atmosphere and can be derived by subtracting ZHD from the TZD value. PWV can be then
calculated as a proportion of the estimated ZWD through the following equation \citep{1987RaSc...22..379A}:
\begin{equation}
                                          PWV =ZWD\times\frac{10^6}{\rho R_V[(k_3/T_m)+k_2]}                                           
\label{eq2}
\end{equation}
where $\rho$ is the density of liquid water, R$_V$ is the specific gas constant for water vapor, k$_2$=70.4 K mbar$^{-1}$,
 k$_3$=3.739 10$^5$ K$^2$mbar$^{-1}$ are the refractivity constants \citep{1994JApMe..33..379B, 1987RaSc...22..379A}, which have
 been determined by direct measurements using microwave techniques (see \cite{1994JApMe..33..379B} for more details on these constants).
 T$_m$ is a weighted mean temperature of the atmosphere, which is strongly correlated with surface temperature (T$_s$) according to the
 linear relation \citep{1992JGR....9715787B}:
  \begin{equation}
                                              T_m = 70.2 + 0.72 T_s    
\label{eq3}
\end{equation}

The uncertainties in the refractivity coefficients of the water vapor and the selected mapping functions in the TZD calculation can introduce a scale factor in the PWV derived from GPS measurements \citep{2001JAtOT..18..830N}. Therefore, the comparison of PWV estimations from GPS data with other techniques will require an initial model consisting of a scale factor between techniques and an offset accounting for a bias in the measurements. This bias comes from residual delays that are estimated as corrections to a delay model assuming a standard atmosphere using a zenith angle dependent mapping function \citep{kleijer01}. The estimation of the PWV from TZD derived by GPS measurements requires surface pressure and temperature measurements. The ZWD will not include an additional bias if accurate surface pressure is used in the estimation of ZHD. Unfortunately, there is not a weather station at the precise location of the GPS antenna at ORM; hence a model \citep{2007JGeod..81..679B}
 of the pressure and temperature at the longitude, latitude and height of the station for each date was used to obtain PWV from the derived TZD using equation \ref{eq1}, \ref{eq2}, and \ref{eq3}. The GPS time series for ORM consist of more than 31500 individual estimations with a
 two-hour temporal resolution of the PWV for the period June 2001 to December 2008. 

During the site prospection for the emplacement of the 10 meters Gran Telescopio de Canarias, different campaigns of PWV measurements above ORM were
 carried out from February 2000 to July 2002 using a two band radiometer. The radiometer operates in the 940-nm water
 vapor band \citep{1998NewAR..42..537K, 2002ASPC..266..220K, 2002ASPC..266..216P, Pinilla03}. Observations were taken on the Moon (solar observations are also possible including an appropiated neutral density filter), measuring the
 entire 940-nm water vapor absorption line (filter centered on 946.7 nm with a bandwidth of 19.8 nm) and nearby continuum (filter centered on 884.5 nm with a bandwidth of 18.2 nm). The PWV is directly correlated to the square of the ratio of the signal in the two filters. No model-dependent assumptions are made for this calculation. The absolute calibration of the 940nm-radiometer was measured from radiosonde flights for a range spanning from 0.5 to 25 mm of water vapor. This remote sensing technique assumes a plane-parallel atmosphere,
 restricting the measurements to nearly Zenithal positions \citep{1989PASP..101..441Q, 1993JApMe..32.1791G} in order to convert the measured line-of-sight PWV to the zenithal values. The relative precision of the line-of-sight PWV is around 5\% for a single measurement \citep{1998NewAR..42..537K}. The two main
 drawbacks of this technique are that is only usable in photometric conditions and with the Moon well above the horizon. 940nm-Radiometer required a human supervision
 and the temporal resolution was about 30 minutes.
 
\begin{figure*}
\centering
\includegraphics[scale=0.40]{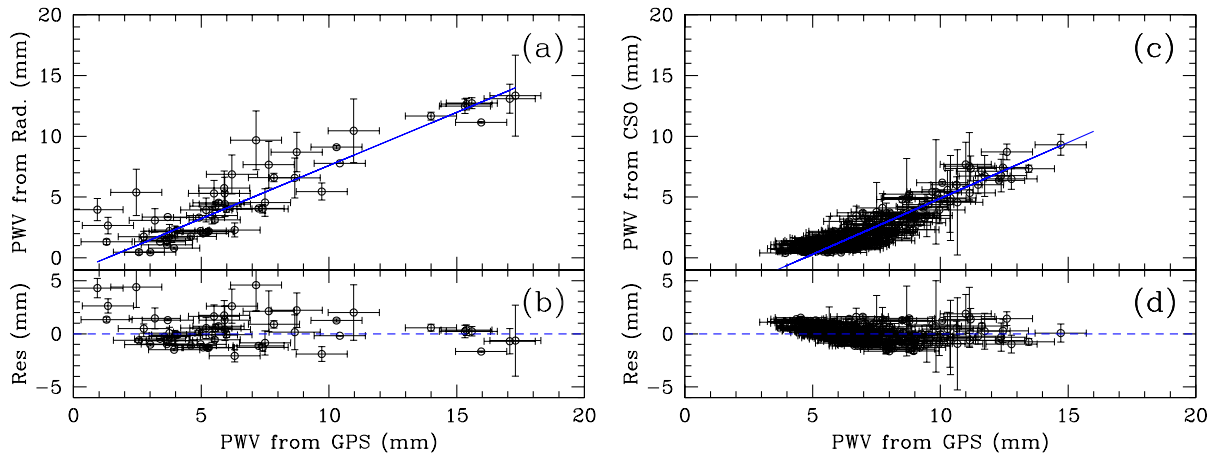}
 \caption{Comparison of PWV estimations from GPS and : (a)-(b) 940nm-radiometer measurements at La Palma site; (c)-(d) PWV estimations from atmospheric opacity at 225 GHz at Mauna Kea site. (a) and (c) The solid line corresponds to the best fit derived from a Bayesian approach to linear regressions with errors in both variables \citep{2007ApJ...665.1489K}. (b) and (d) The residuals of the best-fit model.}
\label{calibra_GPS}
\end{figure*}

A direct comparison between GPS and the 940nm-radiometer could be carried out for the period from June 2001 to July 2002. In this sense, 243 PWV estimations from the 940nm-radiometer were recorded. In order to perform the comparison, we re-sampled the 940nm-radiometer dataset to the same temporal resolution of the GPS data. Following the temporal re-sampling, the number of data was reduced to 55 in both the GPS and the radiometer. As we already mentioned, errors up to 40 \% has been reported for PWV estimations when using radiometer measurements in the 0.5-1 $\mu$m range \citep{1989PASP..101..441Q}. We have adopted the standard deviation of the averaged 940nm-radiometer data in the two-hour sampling as the error of the PWV estimation, being smaller than 25\% for all the estimations. Errors of about 1 mm are estimated for the PWV
 derived from the proccesing of the GPS data. We obtained a good correlation between GPS and 940nm-radiometer measurements of PWV
 for the period June 2001-July 2002 (Fig. \ref{calibra_GPS}), with a Pearson
 correlation coefficient of 0.97. A linear relationship was determined using the task LINMIX\_ERR
 \citep{2007ApJ...665.1489K}, available from the IDL Astronomy User's Library, which incorporates a Bayesian approach to linear
 regressions taking into account measured errors on both variables. The best-fit
 relation and the residuals of the best-fit are shown in Figure \ref{calibra_GPS}. The standard deviation of these residuals is 1.5 mm. The linear regression gives a scale factor of 0.9$\pm$0.1, with an offset at zero of -1.2$\pm$0.1 mm.

\begin{figure*}
\centering
\includegraphics[scale=0.4]{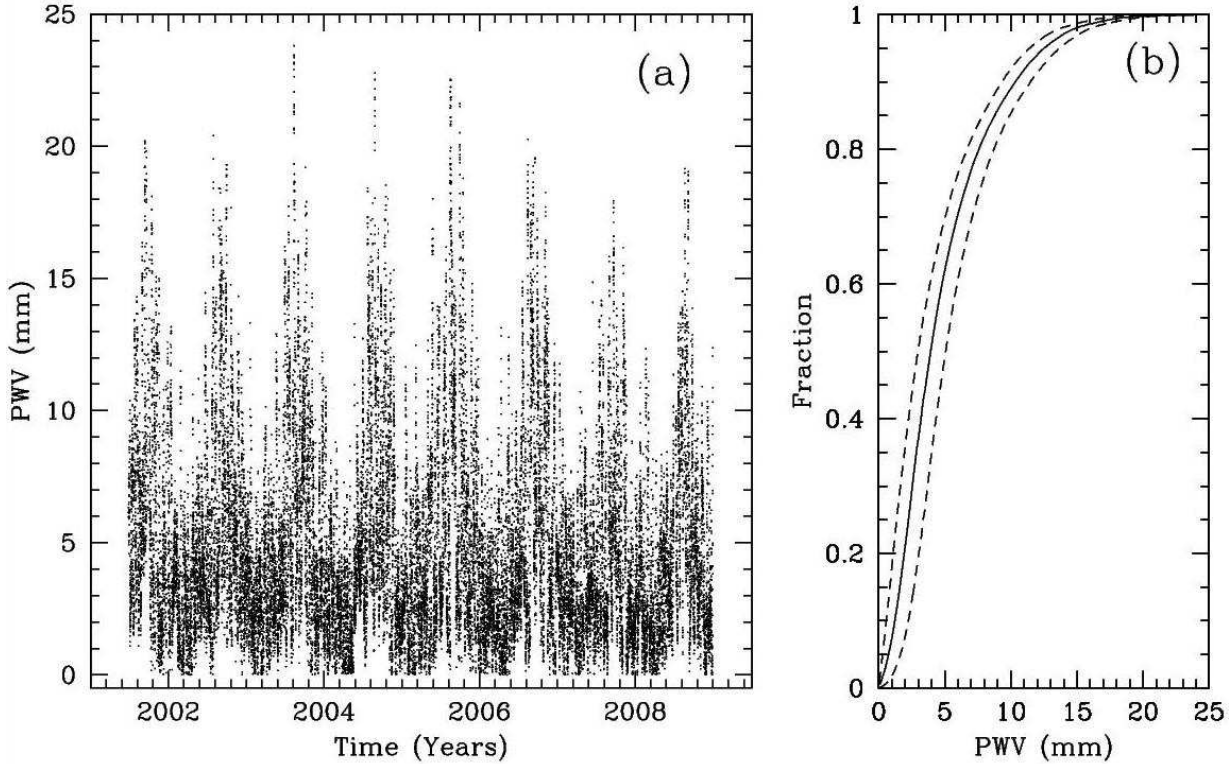}
 \caption{(a) The two-hourly time series of PWV estimations from GPS measurements at ORM from June 2001 to December 2008. (b) Cumulative distribution of PWV measurements (solid line) above ORM. Dotted lines represent the uncertainties due to both the measurements and the calibration.}
\label{serieORM}
\end{figure*}

 The 940nm-radiometer provided the statistical values taken as representative of the ORM site until now. Therefore, we adopted the 940nm-radiometer as the reference instrument to calibrate the long-term GPS time series in order to compensate the scale and offset errors introduced in the data processing. The standard deviation of the residuals was chosen as the uncertainty associated to the PWV estimations from the GPS measurements at ORM. Figure \ref{serieORM} presents the full time series of PWV data derived from GPS measurements above ORM in the period from June 2001 to December 2008. This figure also includes the fraction of measurements for which PWV was found to be below a given value.

Unlike other techniques, GPS provides PWV estimations in nearly all atmospheric conditions, including
 those of local fog, dust and/or presence of clouds (non-photometric conditions). These conditions will induce large PWV
 values that are not considerered in PWV time series obtained from 940nm-radiometer measurements.

\subsection{PWV data at Mauna Kea}


 Mauna Kea is an excellent astronomical site for thermal IR observations. The Caltech
Submillimeter Observatory (CSO) has two monitors (Latitude 19.8224995 North, Longitude -155.475844 West; Altitude: 4070 m above sea level) of
atmospheric opacity, one operating at 225 GHz and another at a wavelength of 350 $\mu$m permanently installed since the
nineties. The characteristics of these radiometers have been extensively described \citep{liu87,mc87, 1994ApOpt..33.1095C}. More information about these monitors is provided in the CSO Tau monitor
webpage (http://puuoo.submm.caltech.edu/). Previous work \citep{1997Icar..130..387D}  has shown that
the atmospheric opacity at 225 GHz, $\tau_{225GHz}$, is related to the column abundance of water vapor in mm, namely:
\begin{equation}
                                             PWV = 20\times(\tau_{225GHz} -0.016)   
\label{eq225_1}
\end{equation}
  
The time series of $\tau_{225GHz}$ measurements
 above Mauna Kea were collected from the CSO archive for the period June 2001 to December 2008. The temporal sampling of $\tau_{225GHz}$
is about 10 minutes. The PWV was estimated (PWV$_{225GHz}$ hereafter) from $\tau_{225GHz}$ data
using equation \ref{eq225_1}. Statistical results based on this reference PWV$_{225GHz}$ time series have been already presented \citep{2009SPIE.7475E..42G, otarola10}.

\begin{figure*}
\centering
\includegraphics[scale=0.40]{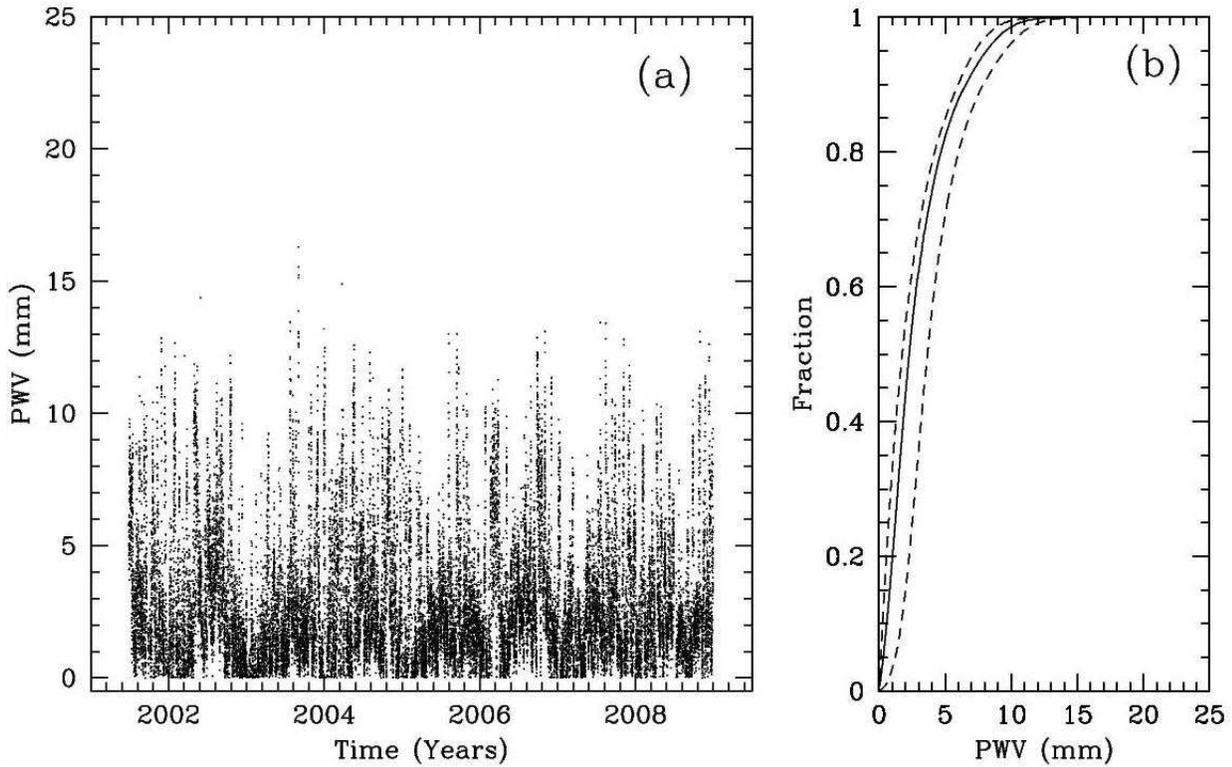}
 \caption{(a) The two-hourly time series of PWV estimations from GPS measurements close to Mauna Kea astronomical site from June 2001 to December 2008. (b) Cumulative distribution of PWV measurements (solid line) above Mauna Kea. Dotted lines represent the uncertainties due to both the measurements and the calibration.}
\label{serieMKEA}
\end{figure*}

A permanent GPS station (labelled MKEA) is located (Latitude:19.8014 North, Longitude:-155.456 West and Altitude: 3755 m above sea level) at approximately
 4 km from Mauna Kea observatory. The coordinate estimation per
process for the GPS antenna at MKEA produces height estimates which are very consistent day to day. This confirms
that the station is correctly installed, stable and that the data is relevant in this
scientific context for water vapor estimations. It is important to point out that
the statistical results from PWV derived using these GPS data might not be representative of the PWV conditions
 at Mauna Kea astronomical site as the GPS station is four km distant from the observatory, although we assume hereafter a similar behavoiur of the PWV in both locations. The difference in altitude between Mauna Kea site and the GPS ($\sim315 m$) might introduce an aditional bias in the PWV estimations derived from GPS relative to the values obtained from $\tau_{225GHz}$. This bias could be up to 1 mm assuming a similar variation of PWV with height as the found in Atacama sites \citep{2001PASP..113..803G} or in Pico Veleta \citep{1989PASP..101..441Q}.
 The data from the MKEA GPS station for the period June 2001-December 2008 were
proccessed using the same global GPS data procedure that we used for the GPS station at La Palma site. In order to compare the GPS time series estimations
 of PWV to PWV$_{225GHz}$, the CSO Tau monitor is adopted as the reference instrument. The PWV$_{225GHz}$ data were re-sampled to the same
 temporal resolution of the GPS time series (2 hours). The standard deviation of the averaged PWV$_{225GHz}$ in the two-hour sampling
is taken as the error of each measurement. Errors of about 1 mm are assumed for the PWV derived from the proccesing of
 the GPS data (PWV$_{GPS}$ hereafter). The number of data to be compared is 20880. The Pearson correlation coefficient
obtained when comparing PWV$_{GPS}$ and PWV$_{225GHz}$ is 0.91. Several random selections of about 500 data were
performed in order to apply the same Bayesian approach to linear regressions than that used for La Palma data. The result of the linear regression
gives a scale factor of 0.9$\pm$0.1 and an intercept of -4.3$\pm$0.1 mm (Fig. \ref{calibra_GPS}). This offset comes from a combination of modeling errors in the GPS treatment and the difference in altitude between the GPS station and the CSO Tau monitor. These scale and offset factors were applied in order to compensate the already explained errors (section \S2.1) introduced in the GPS data processing. The standard deviation of the residuals to the linear
approach is 1.6 mm, which it is assumed hereafter as the uncertanties associated to PWV derived from GPS data at Mauna Kea. The full time series of PWV from GPS measurements obtained for MKEA in the period from June 2001 to December 2008, and the cumulative distribution of the measurements are shown in Figure \ref{serieMKEA}.

\section{Analysis and Results}

GPS measurements provide PWV estimations almost continuously, including daytime and nighttime data. The average PWV at ORM was 4.9 mm, with an standard deviation of 3.7 mm. The
median value considering all the estimations in the PWV time series for ORM is 3.9 mm (Table \ref{PWV_global_stat}). Ephemerides from the Nautical Almanac of
 the Real Observatorio de la Armada de San Fernando (http://www.armada.mde.es) has been used to divide the data in day and nighttime series. The average daytime PWV is 5.1$\pm3.7$ mm, whereas the average nighttime PWV is 4.8$\pm3.7$ mm (Table \ref{PWV_global_stat}).
 The uncertainties indicate only the standard deviation of the averaged measurements. 

The statistical study of PWV variability at different temporal scales is of  relevant importance for an effective scheduling
 at an astronomical observatory with IR capabilities. We have used the time series of PWV
 estimations from GPS data above ORM to study the variability of this
 parameter at seasonal, monthly, and daily scales, but it is important to note that weather patterns are highly
 variable on all timescales. Variability for periods larger than a year (e.g. associated to
 North Atlantic oscillation events and/or a global warming) can not be raised with our current 7.5-year time series.

\begin{table*}
\centering
\caption{Global and seasonal statistical results of precipitable water vapor (in mm) above the Roque de los Muchachos Observatory. The percentage
 of time in which the PWV is within a defined range is also shown. }
\begin{tabular}{|c|c|c|c|c|c|c|c|c|}\hline
           &                     &  {\bf Global}  & {\bf Winter} & {\bf Spring} & {\bf Summer} & {\bf Autumn} \\ \hline
{\bf All:} &                     &                &              &              &              &              \\
           & {\bf Mean}          & 4.9           & 3.3         & 3.5         & 6.7         & 5.6        \\
           & {\bf $\sigma$}      & 3.7           & 2.3         & 2.4         & 4.3         & 3.8        \\
           & {\bf N}             & 30920          & 6886         & 7200         & 8655         & 8179        \\
           & {\bf 10\%}          & 1.2           & 0.9         & 0.9         & 2.0         & 1.4        \\
           & {\bf 25\%}          & 2.3           & 1.7         & 1.8         & 3.2         & 2.6        \\
           & {\bf 50\%}          & 3.9           & 2.9         & 3.0         & 5.7         & 4.8        \\
           & {\bf 75\%}          & 6.7           & 4.4         & 4.7         & 9.4         & 7.9        \\
           & {\bf 90\%}          & 10.3          & 6.3         & 6.6         & 13.0        &  11.13                  \\ \hline
{\bf All:} & {\bf PWV Range}     &               &              &              &              &                \\
           & {\bf $<$ 2 mm}      &   21\%        &      31\%    &    29\%      &  9\%         &   17\%         \\
           & {\bf 2-3 mm}        &   17\%        &      21\%    &    20\%      &  13\%        &   13\%         \\
           & {\bf 3-6 mm}        &   33\%        &      36\%    &    37\%      &  30\%        &   31\%         \\
           & {\bf 6-10 mm}       &   18\%        &      10\%    &    12\%      &  26\%        &   25\%         \\
           & {\bf $\geq$ 10 mm}  &   11\%        &      2 \%    &    2\%       &  22\%        &   14\%         \\ \hline\hline
{\bf Day:} &                     &                &              &              &              &              \\
           & {\bf Mean}          & 5.1           & 3.4         & 3.6         & 6.8         & 5.7        \\
           & {\bf $\sigma$}      & 3.7           & 2.2         & 2.3         & 4.2         & 3.8        \\
           & {\bf N}             & 12967         & 2514         & 3223         & 3904         & 3326        \\
           & {\bf 10\%}          & 1.4           & 1.0         & 1.0         & 2.2         & 1.5        \\
           & {\bf 25\%}          & 2.4           & 1.9         & 2.0         & 3.4         & 2.7        \\
           & {\bf 50\%}          & 4.1           & 3.0         & 3.2         & 5.7         & 4.9        \\
           & {\bf 75\%}          & 6.8           & 4.5         & 4.9         & 9.4         & 8.1        \\
           & {\bf 90\%}          & 10.5          & 6.3         & 6.8         & 12.7        & 11.5       \\ \hline
{\bf Day:} & {\bf PWV Range}                    &               &              &              &              &                \\
           & {\bf $<$ 2 mm}      &   18\%        &      28\%    &    26\%      &  8\%         &   16\%         \\
           & {\bf 2-3 mm}        &   17\%        &      22\%    &    20\%      &  12\%        &   14\%         \\
           & {\bf 3-6 mm}        &   35\%        &      38\%    &    39\%      &  32\%        &   30\%         \\
           & {\bf 6-10 mm}       &   19\%        &      10\%    &    13\%      &  26\%        &   25\%         \\
           & {\bf $\geq$ 10 mm}  &   11\%        &       2\%    &     2\%      &  22\%        &   15\%         \\ \hline\hline

{\bf Night:} &                   &                &              &              &              &              \\
           & {\bf Mean}          & 4.8           & 3.3         & 3.3         & 6.7         & 5.7        \\
           & {\bf $\sigma$}      & 3.7           & 2.3         & 2.4         & 4.4         & 3.7        \\
           & {\bf N}             & 17921         & 4340         & 3977         & 4751         & 4853        \\
           & {\bf 10\%}          & 1.2           & 0.8         & 0.8         & 1.9         & 1.4        \\
           & {\bf 25\%}          & 2.1           & 1.6         & 1.6         & 3.1         & 2.6        \\
           & {\bf 50\%}          & 3.8           & 2.8         & 2.9         & 5.7         & 4.7        \\
           & {\bf 75\%}          & 6.6           & 4.4         & 4.5         & 9.5         & 7.8        \\
           & {\bf 90\%}          & 10.2          & 6.3         & 6.4         & 13.1        & 11.0                  \\ \hline
{\bf Night:} & {\bf PWV Range}                   &               &              &              &              &                \\
           & {\bf $<$ 2 mm}      &   23\%        &      33\%    &     32\%     &  11\%        &   17\%      \\
           & {\bf 2-3 mm}        &   16\%        &      20\%    &     20\%     &  13\%        &   13\%      \\
           & {\bf 3-6 mm}        &   32\%        &      35\%    &     35\%     &  29\%        &   31\%      \\
           & {\bf 6-10 mm}       &   18\%        &      10\%    &     11\%     &  25\%        &   25\%      \\
           & {\bf $\geq$ 10 mm}  &   11\%        &       2\%    &      2\%     &  22\%        &   14\%      \\ \hline\hline

\end{tabular}
\label{PWV_global_stat}
\end{table*}

\subsection{Daily variations}

\begin{figure*}
\centering
\includegraphics[scale=0.50]{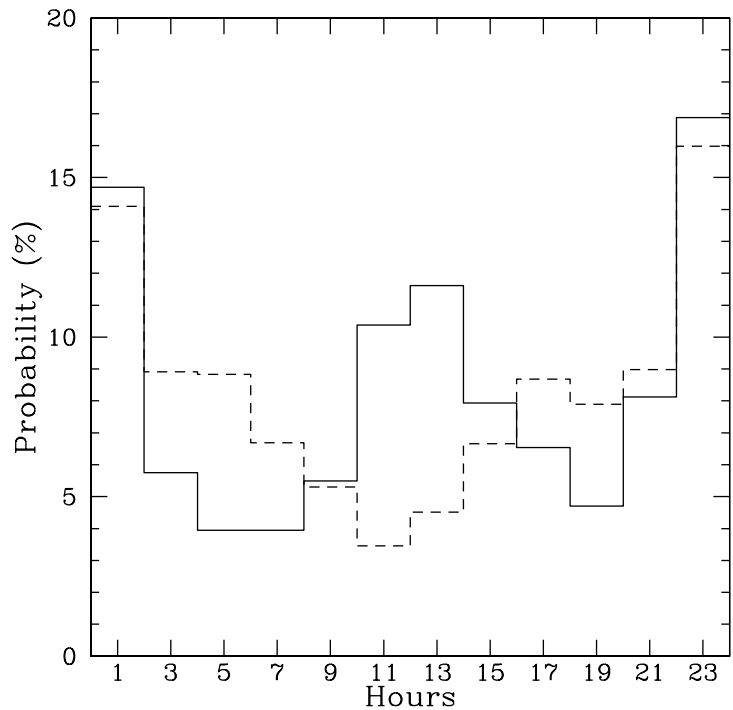}
 \caption{Distribution of hours at which maximum (continuous line) and minimum (dashed line) PWV are reached during a 24-hour period.}
\label{hist_max_min}
\end{figure*}

It is expected that larger temperatures result in larger PWV (as it is expressed in equation \ref{eq2}). Since temperature follows a clear diurnal variation \citep{2002AGUFM.A61C0103Y}, it might be likely that PWV follows a similar trend. The two-hourly time series above La Palma site were used to analyse the variations of PWV in a 24-hour period. The global statistics
 (Table \ref{PWV_global_stat}) suggests a slight difference between 
day and night values of PWV, with larger daytime values. However, such
 differences between daytime and nighttime seems to not follow a clear daily
 behaviour, in the sense that the minimum and maximum PWV during a full day does not have
a defined period. Indeed, maxima or minima values of PWV might appear at any hour during the day as it is illustrated in Fig.
 \ref{hist_max_min}. The distribution of the maximum PWV presents two peaks, 
at around noon and midnight, while the distribution of minimum PWV presents a
 peak only at around midnight. We have tested that this distribution does not change
 significatly if we only take into account the days from a particular season or month. Therefore, a clear daily behaviour of the PWV is not found at ORM.

\subsection{Seasonal variations}

We have derived the global and seasonal statistics of PWV for the period June 2001-December 2008 (see Table \ref{PWV_global_stat}).
 Seasonal differences in the statistical values for daytime and nighttime PWV were found. The smallest statistical PWV
 corresponds to winter and spring, wheareas the largest occurs in summer. Table
\ref{PWV_global_stat} also presents the percentage of time in which PWV lies on a determined range of values defined
 following the criteria to classify the observational conditions for IR observations in \cite{1998NewAR..42..537K}.
 Approximately 38\% of the time (including day and night) the ORM presents PWV values smaller than 3 mm, this percentage increases
 to more than 52\% for winter and spring nighttime. On the contrary, values larger than 6 mm are found 29\% of the
 total time, although this percentage is only about 12\% in winter and spring.

\begin{figure*}
\centering
\includegraphics[scale=0.70]{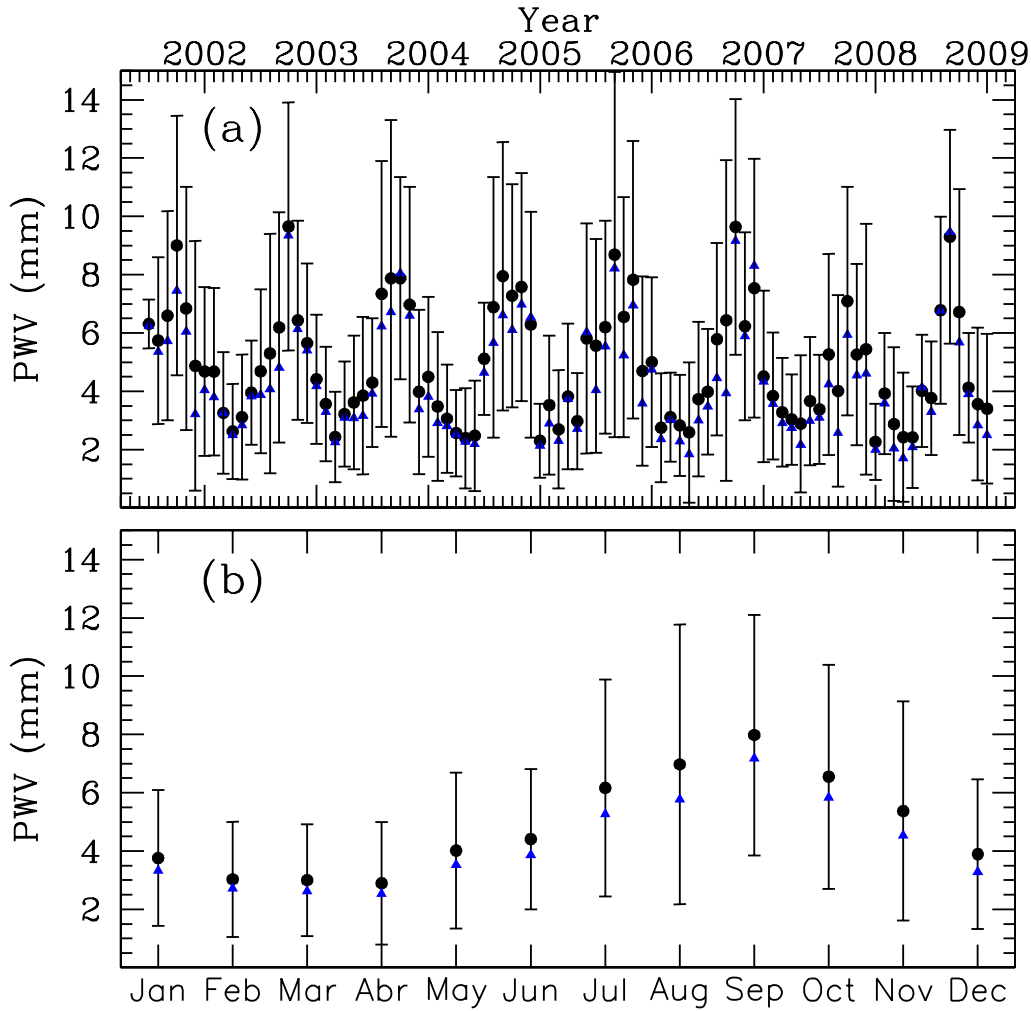}
 \caption{(a) Monthly statistical values of PWV for a 7.5-year period above ORM. This time series constitues a smoothed dataset
 of the two-hour time series derived from GPS measurements recorded at the ORM (including night and day data); (b) The monthly statistics
 of PWV for the period June 2001-December 2008 for ORM. In both plots, dots correspond to the average values and triangles
 to median values. Errorbars only indicate the standard deviation. }
\label{monthly_evolution}
\end{figure*}

We have also derived the monthly average PWV for the different years and for the full period in order to study in detail
the seasonal trend (figure \ref{monthly_evolution}). The PWV presents a clear seasonal behaviour,  being 
September and April the statistically worst and best months, respectively, in terms of PWV. The seasonal behaviour does not significantly differ among the years, showing the largest PWV values during August-September and the lowest during March-April. The latter results were also found if we considered separately the daytime or nighttime series. This seasonal behaviour could be related to the seasonal variation of the temperature (see e.g. the yearly temperature graph for ORM at http://catserver.ing.iac.es/weather/archive/mrtg/temp\_out.html), as the PWV strongly depends on this meteorological variable. Seasonal trends of the PWV have been also reported for other astronomical sites in both hemispheres \citep{otarola10}.

\begin{figure*}
\centering
\includegraphics[scale=0.50]{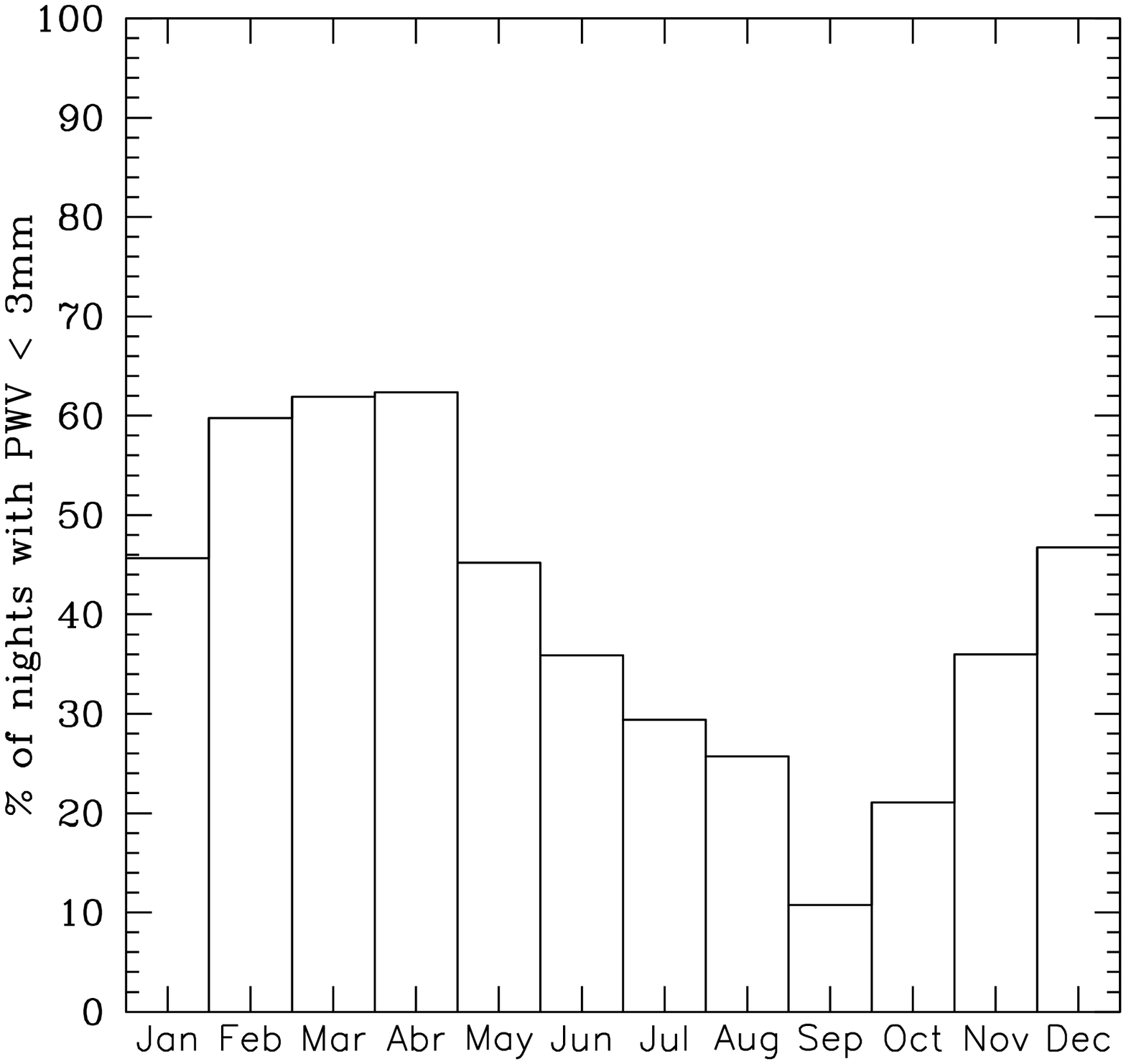}
\caption{Monthly percentage of night time at which PWV $\leq$ 3 mm for the period June 2001-December 2008 at the ORM.  }
\label{per_3mm}
\end{figure*}

If we only take into account the nighttime series, 39\% of the time PWV are lower than 3 mm 
(table \ref{PWV_global_stat}). We have calculated the monthly percentage of night time at which the PWV conditions
 are good or excellent for IR observations, that is PWV $\leq$ 3 mm (Figure \ref{per_3mm}). This percentage also varies along the months. In February, March and April, more than 60\% of the nighttime presents high quality
 conditions to perform IR observations. Even in September, the worst month in terms of PWV, 10\% of the nighttime is highly adequated to perfom IR observations. In this sense, high quality IR observations at ORM could be carried out at any time of the year, being April the best month with $\sim63$\% of the nighttime presenting PWV $\leq$ 3 mm.

\subsection{Temporal Stability}

Temporal stability is an important property in site-testing
 analysis \citep{1996PASP..108..372R, 1997A&AS..125..183M, 2009PASP..121.1151S},
 although it is not usually taken into account. Different astronomical sites with
 similar average values for a particular parameter could strongly differ in the 
temporal stability of that parameter. In this sense, the characterization of the temporal
evolution of a parameter might be
used to optimize the operability of astronomical facilities. The temporal fluctuations of PWV
 drastically affects the schedule of an observing night at a
 telescope/observatory working in queue mode. Therefore, estimations of the range of PWV fluctuations
 on a given time scale is an important factor to take into account for queue-scheduled IR observations.
 The long-time series of PWV estimations recorded at ORM using a two-hour sampling of GPS
measurements allows us to evaluate the temporal variability of the PWV. Variations of PWV
in time scales smaller than two hours can not be studied with the current time series.

In order to study the temporal variability of the PWV at the ORM, we have applied the fractional difference (FD) and the absolute difference (AD) methods. The FD method was introduced by
 \cite{1996PASP..108..372R} to study the temporal
 fluctuations of the seeing at Mauna Kea. The FD method is based on a function
 f$_{FD}$($\Delta$t):
 \begin{equation}
f_{FD}(\Delta t) = <\frac{|PWV(t+\Delta t)-PWV(t)|}{PWV(t+\Delta t)+PWV(t)}>, 
      0\leq f_{FD}(\Delta t) \leq 1
\label{equa2}
\end{equation}

which measures the characteristic range of relative changes as a function of the relative time $\Delta$t. Assuming a log-normal distribution of the data, the characteristic fluctuation time ($\tau_{FD}$) and the growth rate ($\gamma_{FD}$) can be determined by fitting the observations to the following function:

\begin{equation}
<f_{FD}(\Delta t)> = < f_{FD}(\infty)>\times[1- exp(-(\Delta t/\tau)^\gamma)]
\label{equation}
\end{equation}

\begin{figure*}
\centering
\includegraphics[scale=0.35]{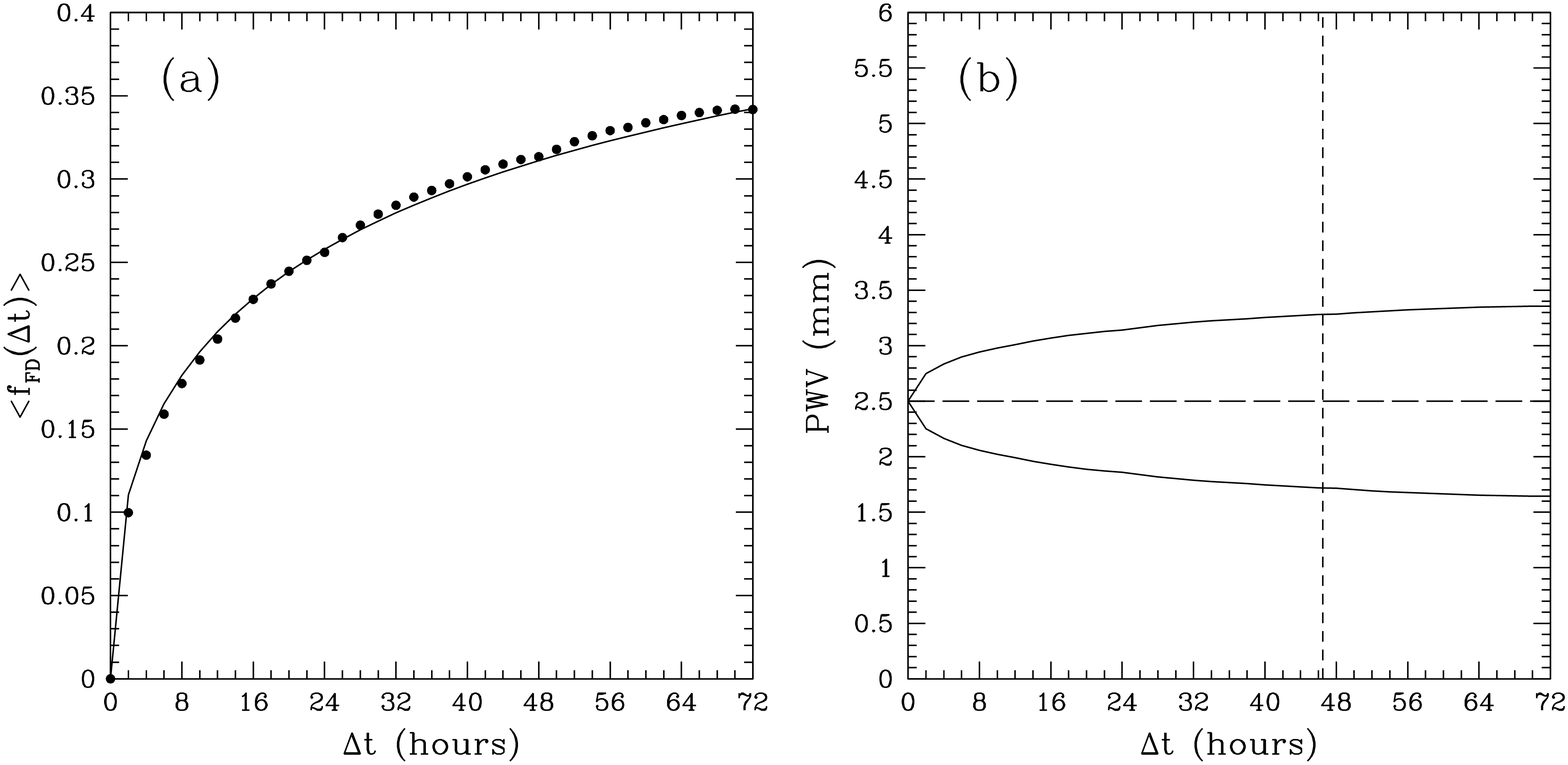}
\caption{(a) Fractional difference of the PWV measurements as a function of the time interval averaged over the full data set. The standard deviation of the $<$f$_{FD}$($\Delta$ t)$>$ is 0.22 in average. (b) PWV range expected (continua lines) at t+$\Delta t$ when assuming a PWV(t)=2.5 mm (long-dashed line) applying the FD method. The short-dashed line indicates the characteristic time $\tau$.   }
\label{pinta_FD}
\end{figure*}

The $<$f$_{FD}$($\Delta t$)$>$ should range from $<$f$_{FD}$(0)$>$=0 to a saturated value $<$f$_{FD}$($\infty$)$>$. From the PWV time series for the ORM, we have computed the average normalized difference for the complete data set for $\Delta t$ ranging from 0 to 72 hours in steps of 2 hours following equation \ref{equa2} (see Fig. \ref{pinta_FD}(a)). The characteristic time and the growth rate ($\tau$ and $\gamma$) have been determined by fitting the equation \ref{equation}
 to $<$f$_{FD}$($\Delta t$)$>$ using a gradient-expansion algorithm (task CURVEFIT from the IDL Astronomy User's
 Library) and minimizing $\chi^2$. The derived values are: $\tau=46.28$ hours, $\gamma=0.43$, and
 $<$ f$_{FD}$($\infty$)$>$ =0.49 mm. The PWV at a time t+$\Delta t$ will be, on average, within the range: PWV(t)$\times$[1$\pm$f$_{FD}$($\Delta t$)]. Let us assume that PWV=2.5 mm at a given time t, the PWV expected at a later time t$\pm\Delta t$ (with $\Delta t$ ranging from 0 to 72 hours) will be in the range between the two curves in figure \ref{pinta_FD}(b).

\begin{figure*}
\centering
\includegraphics[scale=0.35]{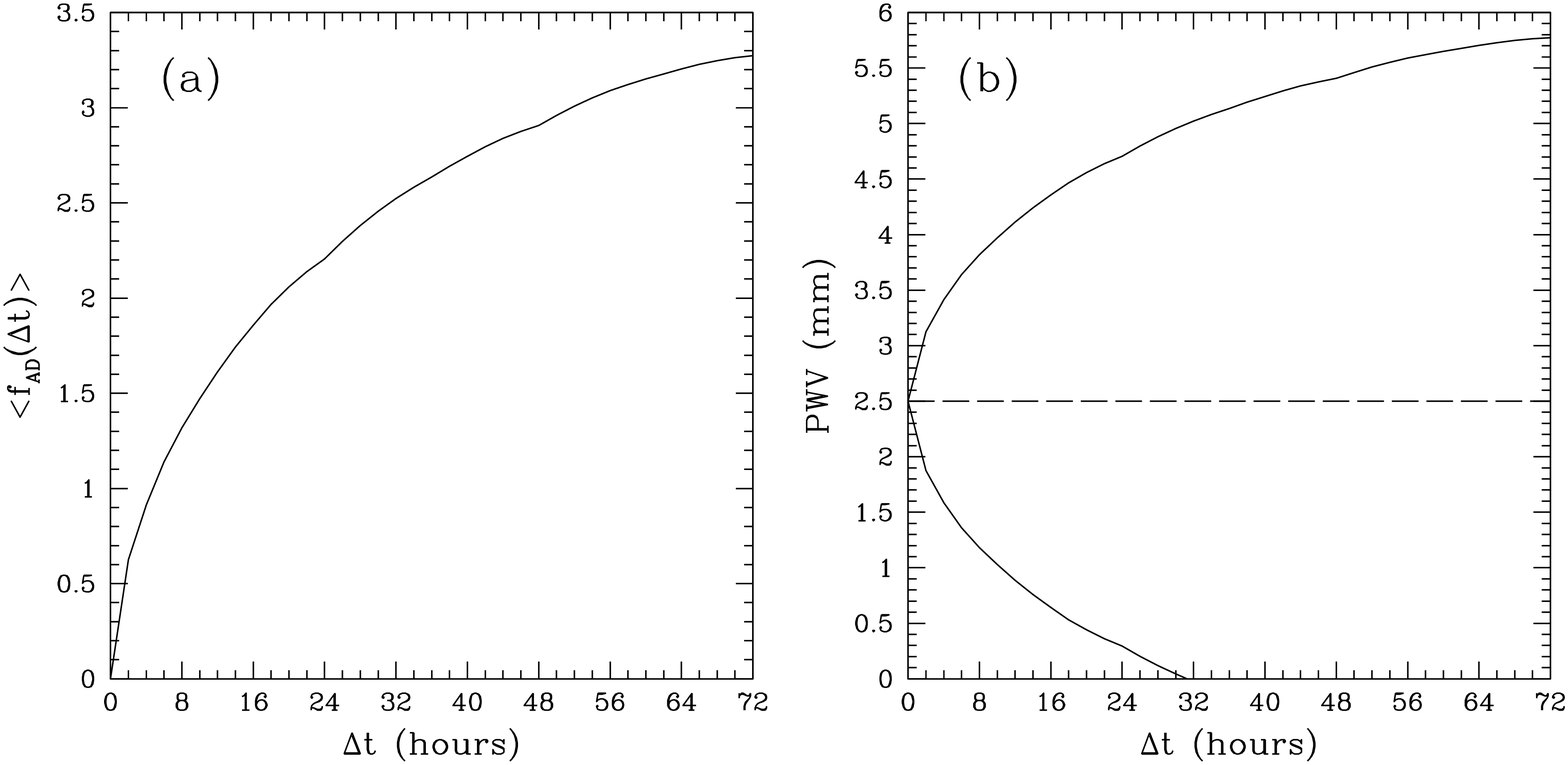}
\caption{(a) Absolute difference of the PWV measurements as a function of the time interval averaged over the full data set. The average standard deviation of the data is 2.30 mm. (b) Expected range of PWV at any time t+$\Delta t$ assuming a PWV equal to 2.5 mm (dashed line) at time t using the AD approach.  }
\label{pinta_AD}
\end{figure*}

The AD method was applied by \cite{2009PASP..121.1151S} to study the temporal variability of the seeing at the five candidate sites for the Thirty Meter Telescope (TMT). The AD method provides information of the absolute range of expected PWV at a time t+$\Delta t$ knowing the PWV at a given time t. The AD is calculated over the entire data set simply applying:
 \begin{equation}
f_{AD} (\Delta t) = <|PWV(t+\Delta t) - PWV(t)|>
\label{equaAD}
\end{equation}
The average $<$f$_{AD}$ ($\Delta t$)$>$ over the entire data set derived from the PWV time series for the ORM is shown in figure \ref{pinta_AD}(a). Let us again assumed that PWV(t)=2.5 mm, the PWV at a time t+$\Delta t$ should be in the range between the two curves in figure \ref{pinta_AD}(b).

According to figures \ref{pinta_FD}(b) and \ref{pinta_AD}(b), and considering $\Delta t$=24 hours, the PWV(t+24) should be in the range 1.86-3.14 mm using the FD approach or in the range 0.29-4.70 mm when the AD method is prefered. The AD method seems to be more restrictive than the FD method, but the reader should take into account that both procedures are averaging over a large number of data and the dispertion is large in both cases.

\begin{figure*}
\centering
\includegraphics[scale=0.35]{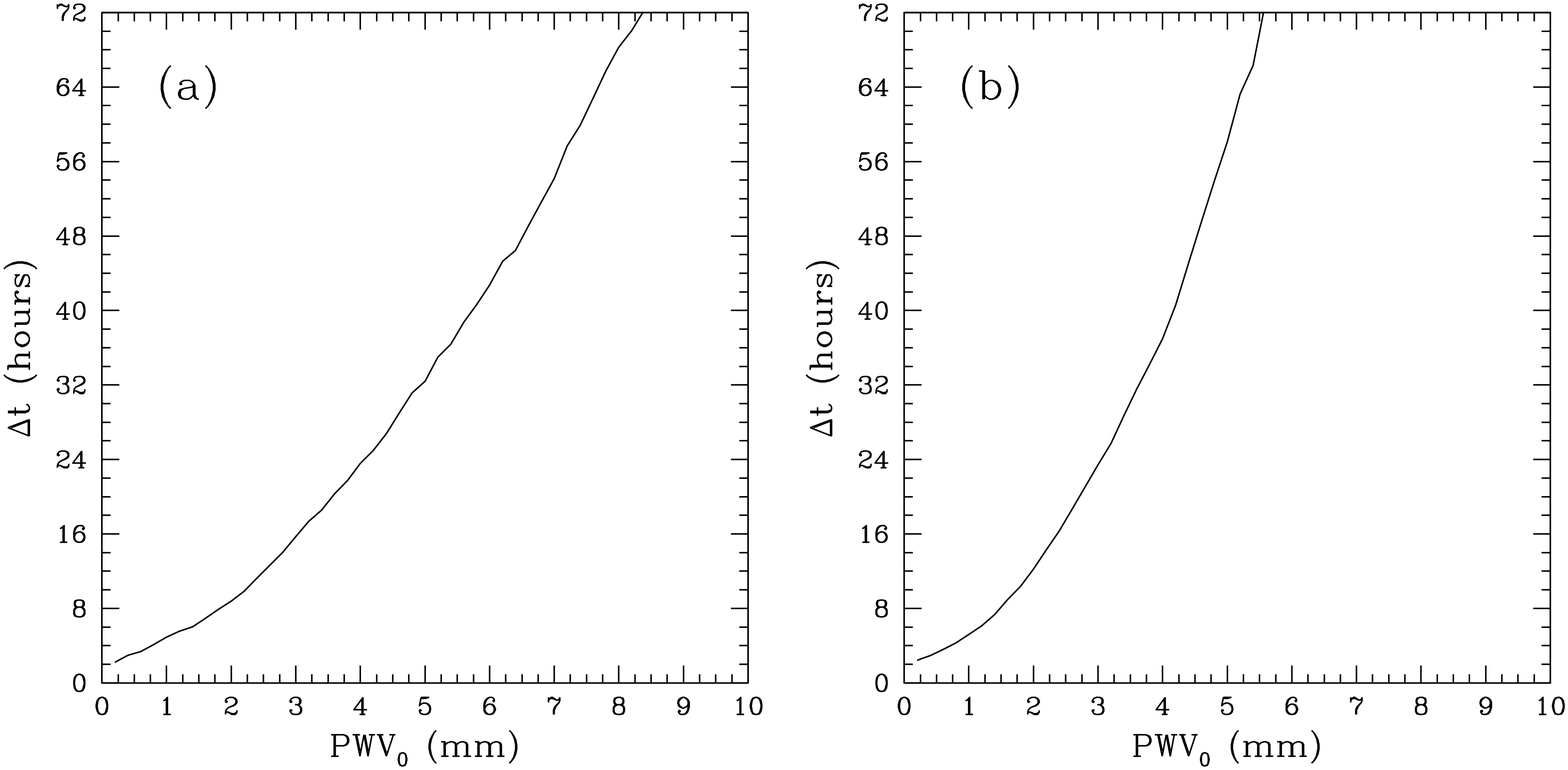}
\caption{(a) Average temporal interval in which the PWV is lower than a defined PWV$_0$ value at ORM; (b) Average temporal interval in which the PWV is lower than a defined PWV$_0$ value at Mauna Kea.  }
\label{pinta_intervalo_medio}
\end{figure*}

We have also calculated the average temporal range in which the PWV is always below a defined value PWV$_0$ (Fig. \ref{pinta_intervalo_medio}). We obtained that the average temporal range in which conditions are good or excellent for IR observations (PWV$\leq$ 3 mm) at the ORM is around 16.9 hours. 

The three methods that were applied to study the temporal variability of the PWV represent statistical analysis of such temporal variations. A steep rise of the PWV in a short time scale can not be ruled out. Indeed, we found that differences larger than 1 mm in consecutive (two hours sampling) PWV measurements occur in $\sim$9\% of the cases. This percentage decreases to $\sim$1.5\% for differences larger than 2 mm in consecutive measurements. A significant improvement in PWV might also be possible. In the two-hour PWV time series for ORM, improvements larger than 1 mm or 2mm also occurs in $\sim$9\% and $\sim$1.6\% of the cases, respectively.

\subsection{PWV statistics at Mauna Kea derived from GPS data}

In order to compare the statistical results on PWV derived for ORM with a higher altitude site, we also carried out
an statistical analysis of PWV estimations derived from GPS measurement at a location closer to Mauna Kea observatory. Previous statistical results
 of PWV have been already obtained from other techniques (see e.g. \cite{2009arXiv0904.3943S,2009SPIE.7475E..42G}) .
 In section \S\ref{compara} we will summarize PWV statistics for different
 worldwide astronomical sites obtained from data recorded using different
 techniques.

\begin{figure*}
\centering
\includegraphics[scale=0.4]{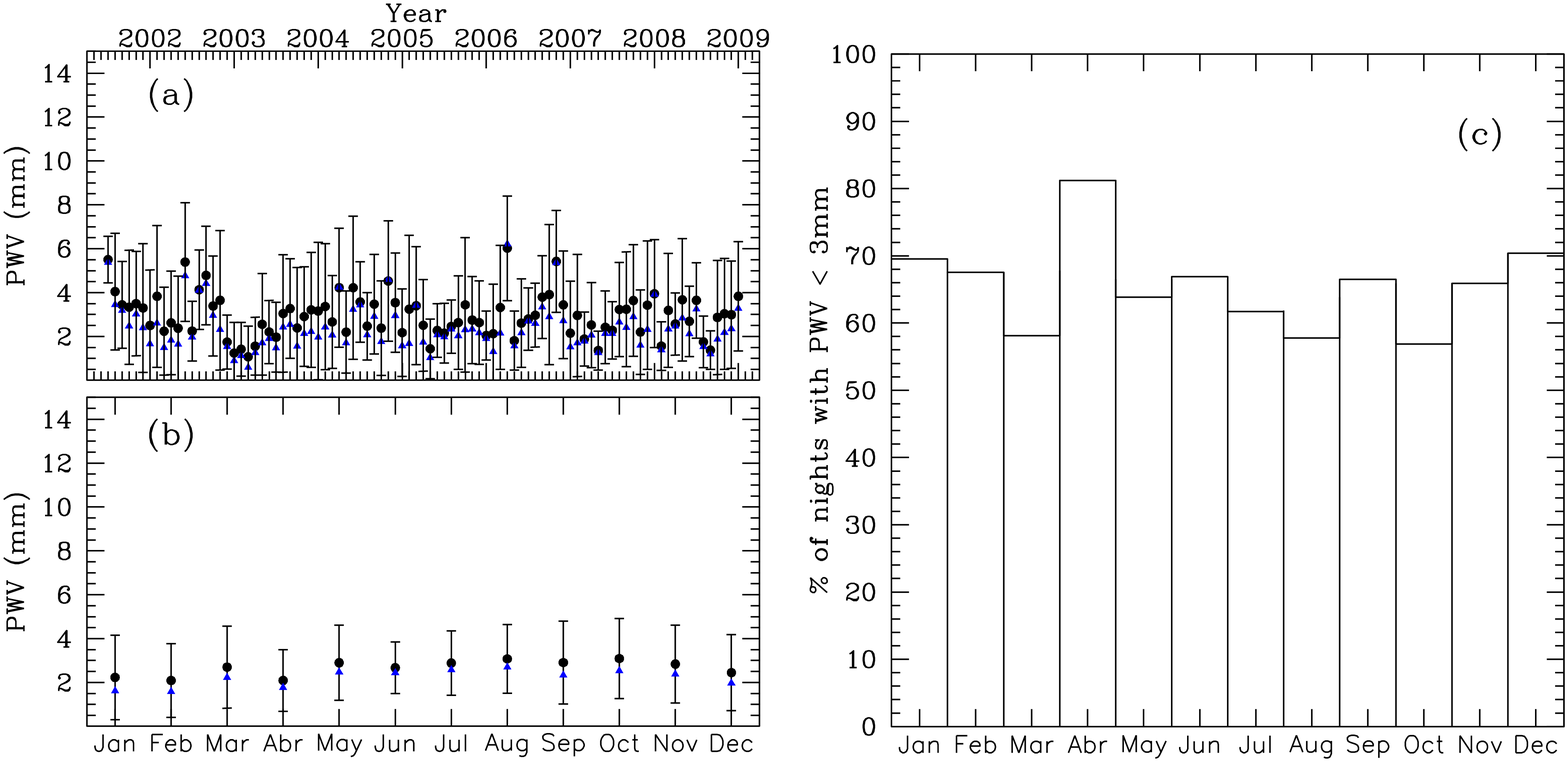}
\caption{(a) Monthly average PWV values for a 7.5-year period above Mauna Kea. This time series constitues a smoothed dataset
 of the two-hour time series derived from GPS data close to Mauna Kea Observatory (including night and day data); (b) The monthly statistics
 of PWV for the period June 2001-December 2008 for Mauna Kea. In both plots, dots correspond to the average values and triangles
 to median values. Errorbars only indicate the standard deviation of the data; (c) Monthly percentage of night time at which PWV $\leq$ 3 mm at Mauna Kea 
for the period from June 2001 to December 2008.  }
\label{figure_MKEA}
\end{figure*}

Table \ref{PWV_global_stat_MKEA} presents the statistical results of the PWV time series derived
from GPS measurements for the period from June 2001 to December 2008 at four km distant from Mauna Kea observatory.
The global average PWV at Mauna Kea is almost 2 mm smaller than the corresponding mean PWV at ORM, which is not surprising
due to the large difference in altitude between both sites (more than 1300 m). Statistically, Summer and Autumn present slightly larger PWV
at Mauna Kea than Winter and Spring. However, a clear seasonal trend in the PWV behaviour is not clear from figures
\ref{figure_MKEA}(a) and \ref{figure_MKEA}(b). The PWV is statistically smaller than 3 mm in 63\% of the time (night+day),
being this value almost independent of the season, although during spring nighttime the percentage increases to 70\%. PWV measurements larger than 6 mm
appear around 12\% of the time at any season. The average temporal range in which conditions are good or excellent for IR observations (PWV$\leq$ 3 mm) at Mauna Kea is 23.4 hours (see figure \ref{pinta_intervalo_medio}). 

Comparing tables \ref{PWV_global_stat} and \ref{PWV_global_stat_MKEA}, we can deduce that for each 10 hours of usefull nighttime
for IR observations (PWV $\leq$ 3mm) at Mauna Kea, the ORM will present 6 useful hours. During Summer and Autumn, only 3.9 and 4.7 hours
 are useful for IR observations at ORM for each 10 useful hours at Mauna Kea, whereas during Winter and Spring these figures increase
to 7.9 and 7.4 useful hours, respectively, for each 10 hours at Mauna Kea. Fair or mediocre conditions for IR observations (3mm $\le$ PWV $\le$ 6 mm)
differ in around 10\% when comparing results from Mauna Kea and ORM, being in any case larger at ORM. During Winter and Spring, ORM presents approximately the same
 percentage of bad conditions (PWV $\geq$ 10 mm) than the percentage at Mauna Kea along the seasons. Moreover, the average temporal range in which conditions are stably excellente or good (PWV$\leq$ 3 mm) are similar (within uncertainties) at Mauna Kea and ORM.

\begin{table*}
\centering
\caption{Global and seasonal statistical results of precipitable water vapor (in mm) above Mauna Kea region including the percentage of time presenting a PWV value in a determined range.}
\begin{tabular}{|c|c|c|c|c|c|c|c|c|}\hline
           &                     &  {\bf Global}  & {\bf Winter} & {\bf Spring} & {\bf Summer} & {\bf Autumn} \\ \hline
{\bf All:} &                     &                &              &              &              &              \\
           & {\bf Mean}          &  3.0        & 2.8       & 2.7      & 3.1       & 3.2      \\
           & {\bf $\sigma$}      &  2.4        & 2.50       & 2.2      & 2.2       & 2.6      \\
           & {\bf N}             &  28607       & 5541       & 6897      & 8417       & 7752      \\
           & {\bf 10\%}          &  0.6        & 0.3       & 0.5      & 0.8       & 0.6      \\
           & {\bf 25\%}          &  1.2        & 0.9       & 1.2      & 1.5       & 1.3      \\
           & {\bf 50\%}          &  2.3        & 1.9       & 2.1      & 2.6       & 2.3      \\
           & {\bf 75\%}          &  4.1        & 4.1       & 3.7      & 4.2       & 4.3      \\
           & {\bf 90\%}          &  6.5        & 6.7       & 5.9      & 6.2       & 7.4      \\ \hline
{\bf All:} & {\bf PWV Range}     &              &            &           &            &                \\
           & {\bf $<$ 2 mm}      &  44\%     & 52\% & 47\% & 36\% & 43\%  \\
           & {\bf 2-3 mm}        &  19\%     & 13\% & 19\% & 21\% & 19\% \\
           & {\bf 3-6 mm}        &  25\%     & 22\% & 24\% & 32\% & 23\% \\
           & {\bf 6-10 mm}       &  11\%     & 12\% & 9\%  & 10\% & 13\% \\
           & {\bf $\geq$ 10 mm}  &  1\%      & 1\%  & 1\%  & 1\%  & 2\%  \\ \hline\hline
{\bf Day:} &                     &          &        &       &        &        \\
           & {\bf Mean}          &  3.1        & 2.8       & 2.8      & 3.3       & 3.3      \\
           & {\bf $\sigma$}      &  2.4        & 2.5       & 2.2      & 2.2       & 2.6      \\
           & {\bf N}             &  16562       & 3072       & 4418      & 5141       & 3931         \\
           & {\bf 10\%}          &  0.7        & 0.4       & 0.7      & 0.9       & 0.7      \\
           & {\bf 25\%}          &  1.3        & 0.9       & 1.3      & 1.6       & 1.3      \\
           & {\bf 50\%}          &  2.4        & 2.0       & 2.2      & 2.7       & 2.4      \\
           & {\bf 75\%}          &  4.2        & 4.1       & 3.8      & 4.3       & 4.5      \\
           & {\bf 90\%}          &  6.6        & 6.7       & 5.9      & 6.3       & 7.5      \\ \hline
{\bf Day:} & {\bf PWV Range}     &              &            &           &            &                \\
           & {\bf $<$ 2 mm}      &  41\%     & 50\% & 44\%  & 33\% & 41\% \\
           & {\bf 2-3 mm}        &  19\%     & 14\% & 21\%  & 22\% & 19\% \\
           & {\bf 3-6 mm}        &  27\%     & 23\% & 25\%  & 34\% & 24\% \\
           & {\bf 6-10 mm}       &  11\%     & 12\% & 9\%   & 10\% & 14\% \\
           & {\bf $\geq$ 10 mm}  &  2\%      & 1\%  & 1\%   & 1\%  & 2\%  \\ \hline\hline

{\bf Night:} &                   &          &        &       &        &        \\
           & {\bf Mean}          & 2.9         & 2.7       & 2.5      & 2.9       & 3.1      \\
           & {\bf $\sigma$}      & 2.4         & 2.5       & 2.3      & 2.2       & 2.6      \\
           & {\bf N}             & 12043        & 2462       & 2484      & 3276       & 3821         \\
           & {\bf 10\%}          & 0.5         & 0.3       & 0.4      & 0.7       & 0.6      \\
           & {\bf 25\%}          & 1.1         & 0.8       & 0.9      & 1.4       & 1.2      \\
           & {\bf 50\%}          & 2.1         & 1.8       & 1.8      & 2.4       & 2.2      \\
           & {\bf 75\%}          & 3.9         & 3.9       & 3.5      & 4.0       & 4.1      \\
           & {\bf 90\%}          & 6.5         & 6.7       & 5.7      & 5.9       & 7.3                 \\ \hline
{\bf Night} & {\bf PWV Range}     &              &            &           &            &                \\
           & {\bf $<$ 2 mm}      & 48\%   & 55\% &  53\%  & 42\% & 45\% \\
           & {\bf 2-3 mm}        & 17\%   & 12\% &  17\%  & 19\% & 19\% \\
           & {\bf 3-6 mm}        & 23\%   & 20\% &  21\%  & 29\% & 22\% \\
           & {\bf 6-10 mm}       & 10\%   & 12\% &  8\%   & 9\%  & 12\% \\
           & {\bf $\geq$ 10 mm}  & 2\%    & 1\%  &  1\%   & 1\%  & 2\%  \\ \hline\hline

\end{tabular}
\label{PWV_global_stat_MKEA}
\end{table*}

This comparison demostrates that excellent conditions for IR astronomical observations, in terms of percentage and stability, are also possible in sites at 
relative low altitude (over 2000 m above sea level). This result might be related to the role of the troposphere thickness instead of the site altitude
in the IR quality of a particular astronomical site \citep{2004SPIE.5572..384G}. Indeed, the seasonal behaviour of the PWV
at ORM follows a quite similar behaviour than the troposphere thickness at this site (see figure 5 in \cite{2004SPIE.5572..384G}).

\subsection{Summary of PWV statistical results at astronomical sites}
\label{compara}

The comparison of PWV statistical results between astronomical sites is very complex due to the variety of techniques and procedure normally used.
 Different techniques provide different temporal coverages and samplings. Moreover, some of the techniques used to evaluate the
 water vapor content are strongly affected by rain and clouds (e.g. radiosondes and WVR). Others are only applicable near the zenith (e.g. radiometers), whereas
 GPS can provide continuous PWV estimates depite the effects of rain, dust or clouds. In spite of such differences, we summarize in table \ref{PWV_summarize} 
statistical PWV results for different astronomical sites located at different altitudes above sea level. In the case of Mauna Kea, we have found
median PWV values derived from four different techniques, showing significant discrepancies between the values. Even using the same database but different analysis period, statistical values may differ significantly. These differences also occur
 in the case of ORM, where Infrared sky radiance gives a significant lower PWV median than the other techniques used there. In both sites, 
GPS provides the largest PWV median value, as it is expected from the fact that GPS measures continuosly and under all weather conditions. 
 \begin{table*}
\centering
\caption{Statistical results of precipitable water vapor for different astronomical sites worldwide derived from different techniques and procedures.}
\begin{tabular}{|l|l|c|c|l|c|l|}\hline
{\bf Site} & {\bf Location: } & {\bf Height} & {\bf Median} & {\bf Technique} & {\bf Temporal} & {\bf Reference } \\
           & {\bf Lat, Lon}   & {\bf (m)}    & {\bf PWV (mm)} &               & {\bf Range}          &                  \\ \hline
Las Campanas & 29.01$^\circ$ S, 42.18$^\circ$ W & 2200          &   2.8         & 225GHz-radiometer & 2005/07-08 & \cite{2007PASP..119..697T} \\ 
Cerro Tolar& 21.96$^\circ$ S, 70.10$^\circ$ W  & 2290            &   4.02        & GOES-8 satellite & 1993/06-1996/02 & \cite{otarola10} \\
           &                                  &                &   4.7         & Surface PWV data  & 01/2004-12/2007           &   \cite{otarola10} \\
ORM        & 28.77$^\circ$ N, 17.88$^\circ$ W  & 2395            &   3.9         & 940nm-Radiometer & 1996-1998 & \cite{1998NewAR..42..537K} \\
           &                                  &                 &   3.9          & GPS & 2001/06-2008/12 & this work \\
           &                                  &                 &   2.6        &   IR Sky Radiance & 2000-2002 & \cite{Pinilla03} \\
La Silla   & 29.25$^\circ$ S, 70.73$^\circ$ W  & 2400            &   3.9        & IR Sky Radiance & 1983-1989 & \cite{report} \\
Paranal    & 24.63$^\circ$ S, 70.40$^\circ$ W  & 2635            &   2.3        & IR Sky Radiance & 1983-1989 & \cite{report} \\
San Pedro M\'artir & 31.04$^\circ$ N, 115.47$^\circ$ W  & 2830   &   2.63       & GOES-8 satellite & 1993/06-1996/02 & \cite{otarola10} \\
                   &                                  &        &    3.4       & 210GHz-radiometer    & 2006           &   \cite{otarola10} \\
Pico Veleta& 37.07$^\circ$ N, 3.37$^\circ$ W   & 2850            &   2.9          & 940nm-Radiometer & 1984-1987 & \cite{1989PASP..101..441Q} \\
Cerro Armazones & 24.58$^\circ$ S, 70.18$^\circ$ W   & 3064      &   2.87         &  GOES-8 satellite & 1993/06-1996/02 & \cite{otarola10} \\
                &                                  &            &   3.2         & Surface PWV data  &    01/2004-12/2007        &   \cite{otarola10} \\
Dome C     & 75.06$^\circ$ S, 123.23$^\circ$ E & 3233            &   0.34         & Satellite \& Model &  2008                & \cite{2009PASP..121..976S} \\
Mauna Kea  & 19.83$^\circ$ N, 155.47$^\circ$ W & 4205            &   1.7         & 225GHz-radiometer  & 2001/06-2008/12 & \cite{2009SPIE.7475E..42G}  \\
           &                                  &                 &   1.2         & Radiosondes        & 1983                    & \cite{1987PASP...99..560B}  \\
           &                                  &                 &   1.86        & GOES-8 satellite & 1993/06-1996/02 & \cite{otarola10} \\
           &                                  &                 &   2.1         & 225GHz-radiometer  &  01/2004-12/2007       & \cite{otarola10} \\
           &                                  &                 &   2.3         & GPS                & 2001/06-2008/12 & this work  \\
Ridges A   & 81.5$^\circ$ S, 73.5$^\circ$ E    & 4053            &   0.21         & Satellite \& Model &  2008                & \cite{2009PASP..121..976S} \\
Dome A     & 80.73$^\circ$ S, 77.3$^\circ$ E   & 4083            &   0.23         & Satellite \& Model &  2008                & \cite{2009PASP..121..976S} \\
Cerro Tolonchar & 23.93$^\circ$ S, 67.97$^\circ$ W & 4480        &   1.7          & GOES-8 satellite & 1993/06-1996/02 & \cite{otarola10} \\
                &                                &             &   1.8          & Surface PWV data  &  01/2004-12/2007           &   \cite{otarola10} \\  
Chajnantor & 23.02$^\circ$ S, 67.45$^\circ$ W  & 5080            &   1.0         & Radiosondes         & 1998/10 \& 2000/08 & \cite{ 2001PASP..113..803G} \\
           &                                  &                 &   1.2         & 225GHz-radiometer &  1995/04-2000/04 & \cite{2007PASP..119..697T} \\ \hline

\end{tabular}
\label{PWV_summarize}
\end{table*}

In order to calculate an approximate PWV median value comparable to that derived from the techniques affected by clouds, we have selected the GPS data recorded
 during photometric nights
(clear nights from clouds and/or aerosol affecting astronomical observations). The atmospheric extinction coefficient -- a measurement of the sky transparency --
is measured at the ORM in the V (551 nm, K$_V$ hereafter) and  r' (625 nm) bands since 1984 by the Carlsberg Automatic Meridian Circle (CAMC). We have downloaded the
atmospheric extinction coefficient for the period June 2001-December 2008 from the CAMC archive (http://www.ast.cam.ac.uk/\~dwe/SRF/camc\_extinction.html). We selected
as photometric nights those satisfying the following requirements: (i) the total number of observing hours is larger than seven; (ii) the total number
of photometric data taken is larger than 90\% of the total number of observing hours; and (iii) the average K$_V$ along the night is smaller than 0.2. 55\% of the nights from
 June 2001 to December 2008 in the CAMC database follow the selection criteria. PWV measurements derived from GPS data during photometric nights were then used to derived
 the statistical values of PWV (see table \ref{statistic_photometric}).

\begin{table*}
\centering
\caption{Seasonal statistical results of precipitable water vapor (in mm) above the Roque de los Muchachos Observatory derived from PWV estimations during 
photometric nights.  }
\begin{tabular}{|c|c|c|c|c|c|c|c|c|}\hline
           &                     &  {\bf Global}  & {\bf Winter} & {\bf Spring} & {\bf Summer} & {\bf Autumn} \\ \hline
{\bf Night:} &                     &                &              &              &              &              \\
           & {\bf Mean}          & 3.9           & 3.2         & 3.1         & 4.7         & 4.5        \\
           & {\bf $\sigma$}      & 2.6           & 1.9         & 1.9         & 2.9         & 2.8        \\
           & {\bf N}             & 6146           & 1256         & 1517         & 2081         & 1292        \\
           & {\bf 10\%}          & 1.2           & 0.9         & 0.9         & 1.7         & 1.4        \\
           & {\bf 25\%}          & 2.1           & 1.7         & 1.7         & 2.6         & 2.3        \\
           & {\bf 50\%}          & 3.4           & 2.8         & 2.9         & 3.9         & 4.0        \\
           & {\bf 75\%}          & 5.2           & 4.2         & 4.2         & 6.2         & 6.2        \\
           & {\bf 90\%}          & 7.3          & 5.7         & 5.7        & 8.4       &  8.3                \\ \hline
\end{tabular}
\label{statistic_photometric}
\end{table*}

Statistical values (average and median) of PWV strongly improve when specific weather conditions are selected, which is
usually imposed by the techniques used to retrive PWV values. Therefore, the comparison of PWV values from different sites is only possible when the same
technique and procedure have been used to adquire and analyse the data.  

\subsection{Infrared bands and PWV}

ORM has been an astronomical site traditionally dedicated to optical and near-IR observations (from 0.4 to 2.5 $\mu$m). The 10 meters Gran Telescopio de Canarias will extend this range to the mid-IR, with CanariCam, a camera and spectrograph working in the thermal infrared between $\sim$7.5 and 25 
$\mu$m. In this section, we explore the implications in terms of atmospheric transmission of PWV values in the standard near and mid-IR windows in
 astronomy. We have modelled the theoretical near and mid-IR transmission spectrum for La Palma site using the ATRAN modelling software (Lord,
 S.D. 1992, NASA Technical Memor. 103957) throughout the web-based form at http://atran.sofia.usra.edu/cgi-bin/atran/atran.cgi . We selected the closest
 latitude to the site allowed (30 degrees), the altitude of the site (2400 m), scanning the PWV from 1 to 10 mm. The transmission spectrum
 derived from each PWV value was integrated in the spectral range covering the standard filters windows and these values were refered to the
 integrated transmission asumming a PWV equal to 1 mm (see table \ref{trans}). The theoretical transmission spectrum for Mauna Kea was
 extracted from a model available at the GEMINI Observatory web pages (http://staff.gemini.edu/~kvolk/linkpage.html) which is also based
 on the ATRAN modeling software, although it is limited in PWV values. We followed the some
 procedure than in the case of La Palma, integrating the transmission spectrum in each filter window and refering the value to the derived
 for 1mm of PWV.

\begin{table*}
\centering
\caption{Atmospheric transmission as a function of PWV relative to the atmospheric transmission extracted for 1 mm of PWV for the standard near and mid infrared bands.}
\begin{tabular}{|l|c|c|c|c|c|c|c|c|}\hline
{\bf Near } & {\bf PWV } & {\bf Filter J} & {\bf Filter H} &  {\bf Filter K} & {\bf Filter L} & {\bf Filter M} & {\bf Filter N} & {\bf Filter Q} \\
{\bf La Palma}                     & (mm)      &  (1.1-1.4 $\mu$m) & (1.5-1.8 $\mu$m) & (2.0-2.4 $\mu$m) & (3.0-4.0 $\mu$m) & (4.5-5.1 $\mu$m) & (10-12 $\mu$m) & (17.5-22.5 $\mu$m) \\ \hline
                     &   1.5     &  97.4 &     99.7  &    99.8  &    97.5  &    97.5  &    99.8  &    89.3  \\
                     &   3.0     &  92.5 &     98.8  &    99.1  &    92.2  &    91.8  &    99.4  &    68.2  \\
                     &   5.0     &  88.7 &     98.1  &    98.4  &    87.9  &    87.2  &    99.0  &    54.7  \\
                     &   6.5     &  86.8 &     97.6  &    97.9  &    85.6  &    84.9  &    98.8  &    48.9  \\ 
                     &   7.6     &  85.8 &     97.4  &    97.7  &    84.4  &    83.9  &    98.7  &    47.2  \\ 
                     &   10.0    &  83.2 &     96.7  &    96.9  &    80.9  &    79.4  &    98.3  &    36.5  \\ \hline
{\bf Mauna} & {\bf PWV } & {\bf Filter J} & {\bf Filter H} &  {\bf Filter K} & {\bf Filter L} & {\bf Filter M} & {\bf Filter N} & {\bf Filter Q} \\
{\bf Kea}                & (mm)      &  (1.1-1.4 $\mu$m) & (1.5-1.8 $\mu$m) & (2.0-2.4 $\mu$m) & (3.0-4.0 $\mu$m) & (4.5-5.1 $\mu$m) & (10-12 $\mu$m) & (17.5-22.5 $\mu$m) \\ \hline
                     &   1.5     &  98.6 &     99.8  &    99.9  &    99.0  &    98.8  &    99.1  &    84.8  \\
                     &   3.0     &  91.1 &     98.7  &    99.1  &    93.3  &    92.2  &    98.2  &    68.3  \\
                     &   5.0     &  87.2 &     97.9  &    98.4  &    89.7  &    88.1  &    96.9  &    55.9  \\
                     &   6.5     &  86.1 &     97.6  &    98.6  &    89.0  &    88.0  &     ---  &    --- \\ \hline
\end{tabular}
\label{trans}
\end{table*}

In both cases, similar results were found. The atmospheric transmission in bands H, K, and N decrease less than 5\% for PWV variations from 1 to 10 mm. The atmospheric transmission in windows J, L, and M slowly decrease when PWV increase (up to 21\% from 1 to 10 mm).
 The most significant effect occurs in band Q, where atmospheric transmission is around 50\% smaller when PWV increase from 1 to 5 mm. When PWV is 10 mm, the atmospheric transmission is almost 35\% of the transmission with 1 mm of PWV in Q-band. It is also important to note that high levels of water vapor also increase the thermal IR background, which is one of the major factors limiting the atmospheric transparency in mid-IR observations from ground-based sites. 
The effects of observing conditions (water vapor column, airmass, cloud cover, etc) in the quality of data recorded in mid-IR bands was illustrated in \cite{2008SPIE.7016E..63M} using real data obtained in Mauna Kea.

\section{Summary and Conclusions}

We have analised the water vapor content for the period from June 2001 to December 2008 above the ORM using PWV estimations from GPS data. We have verified the consistency of 940nm-radiometer and GPS estimation of PWV, removing the offset between both techniques. We have also presented statitical results for Mauna Kea for the same period, analysing the GPS close to Mauna Kea site with $\tau_{225GHz}$ measurements. Our main results and conclusions may be summarized as follows:

\begin{itemize}

\item[(i)] The nighttime median PWV above ORM is 3.79 mm, with slightly differences between day and nighttime statistics.

\item[(ii)] The PWV presents a clear seasonal variation at the ORM. Winter and spring nights present the lower PWV statistics, with a median value of 2.82 mm and 2.86 mm, respectively.

\item[(iii)] More than 60\% of the nighttime during February, March and April present good or excellent conditions in terms of water vapor (PWV $\leq$ 3 mm) at the ORM.

\item[(iv)] Comparing PWV statistical results for ORM and Mauna Kea, we deduce that for 10 hours of good conditions for IR observations at Mauna Kea during winter, the ORM will present 7.9 hours showing excellent conditions.

\item[(v)] The average temporal range presenting good or excellent conditions for IR observations (PWV $\leq$ 3mm) is comparable at ORM and Mauna Kea (16.9 hours and 23.4 hours, respectively).

\item[(vi)] The comparison of PWV at different astronomical sites is difficult because there is not an unique defined technique/procedure to estimate the PWV.

\end{itemize}

GPS is a promising technique to unify the PWV estimations at many astronomical sites.

\section*{Acknowledgments}

This paper is based on GPS data recor
ded from the GPS station at the
 Roque de los Muchachos Observatory on the island of La Palma and from the
 labelled MKEA station at Mauna Kea on the island of Hawaii. Both GPS stations are part of the international network of GPS stations EUREF (www.euref.eu).

 Measurements of the 225 GHz optical depths at Mauna Kea were
 recorded from the web page of the Caltech Submillimeter Observatory (CSO)
(http://puuoo.submm.caltech.edu/ ). The coefficient-extinction data were
 downloaded from the database (http://www.ast.cam.ac.uk/\~ dwe/SRF/camc\_extinction.html) of the Carlsberg Meridian Circle of the Isaac Newton Group on. 

Thanks are due to M. Kidger who managed the water vapor monitoring campaigns
at the ORM from 1996 to 2002. The authors also thank I. Romero from Canary Islands Advanced Solutions ( http://www.sacsl.es/) and J.A. Acosta-Pulido from Instituto de Astrof\'{\i}sica de Canarias for stimulating discussions and help. Finally, we wish to thank the referee for very constructive comments and suggestions.

This work was partially funded by the Instituto de Astrof\'{\i}sica de Canarias
 and by the Spanish Ministerio de Educaci\'on y Ciencia (AYA2006-13682 and AYA2009-12903). This
 work has been carried out within the framework of the European Project OPTICON
 and under Proposal FP6 for Site Selection for the European ELT. 
B. Garc\'{\i}a-Lorenzo and A. Eff-Darwich thank the support
from the Ram\'on y Cajal program by the Spanish Ministerio de Educaci\'on y
 Ciencia.

\end{document}